\documentclass[conference,compsoc]{IEEEtran}
\IEEEoverridecommandlockouts

\usepackage{amsmath,amssymb,amsfonts}
\usepackage{algorithmic}
\usepackage{graphicx}
\usepackage{textcomp}
\usepackage[dvipsnames]{xcolor}
\def\BibTeX{{\rm B\kern-.05em{\sc i\kern-.025em b}\kern-.08em
    T\kern-.1667em\lower.7ex\hbox{E}\kern-.125emX}}
\usepackage[backend=biber,style=numeric,maxnames=99,minnames=99]{biblatex}
\usepackage[inline]{enumitem}
\usepackage{xspace}
\usepackage[binary-units]{siunitx}
\usepackage[]{hyperref} 
\usepackage{adjustbox}
\usepackage{supertabular, multicol}
\usepackage{booktabs}
\usepackage{colortbl}
\usepackage{listings}
\usepackage{tabularx}
\usepackage{multirow}

\usepackage{subcaption}\usepackage{pifont}
\newcommand{\cmark}{{\color{Green}\ding{51}}}
\newcommand{\xmark}{{\color{Red}\ding{55}}}

\graphicspath{{./fig/}}
\DeclareGraphicsExtensions{.pdf}

\definecolor{mygreen}{rgb}{0,0.6,0}
\definecolor{mygray}{rgb}{0.5,0.5,0.5}
\definecolor{mymauve}{rgb}{0.58,0,0.82}
\colorlet{rowgray}{gray!15}
\colorlet{hlgreen}{Green!25}

\usepackage{array}
\newcolumntype{C}[1]{>{\centering\arraybackslash}m{#1}}

\DeclareSIUnit{\cpuhour}{CPU\textrm{-}hours}
\DeclareSIUnit{\loc}{LOC}

\ExplSyntaxOn
  \cs_new_eq:NN \calc \fp_eval:n
\ExplSyntaxOff

\def\sys{Magma\xspace}
\def\sysurl{\url{https://hexhive.epfl.ch/magma/}\xspace}
\def\lavam{LAVA-M\xspace}

\def\libpngbugn{7}
\def\libtiffbugn{14}
\def\libxmlbugn{18}
\def\popplerbugn{22}
\def\opensslbugn{21}
\def\sqlitebugn{20}
\def\phpbugn{16}
\def\systargetcount{seven\xspace} 
\def\sysdrivercount{25\xspace}
\def\sysbugcount{\calc{\libpngbugn+\libtiffbugn+\libxmlbugn+\popplerbugn+\opensslbugn+\sqlitebugn+\phpbugn}\xspace} 
\def\syscwecount{11\xspace}

\newlist{inlineroman}{enumerate*}{1}
\setlist[inlineroman]{label=(\roman*)}

\newlist{inlinealph}{enumerate*}{1}
\setlist[inlinealph]{label=(\alph*)}

\begin{document}

\title{The Impact of Magma: A Ground-Truth Fuzzing Benchmark}

\makeatletter
\newcommand{\linebreakand}{%
  \end{@IEEEauthorhalign}
  \hfill\mbox{}\par
  \mbox{}\hfill\begin{@IEEEauthorhalign}
}
\makeatother

\author{
\IEEEauthorblockN{Ahmad Hazimeh*
\thanks{*Work completed prior to the author joining BugScale.}}
\IEEEauthorblockA{\textit{EPFL; BugScale} \\
ahmad@hazimeh.dev}
\and
\IEEEauthorblockN{Adrian Herrera\dag{}
\thanks{\dag{}Work completed prior to the author joining Interrupt Labs.}}
\IEEEauthorblockA{\textit{Australian National University; Interrupt Labs} \\
adrian.herrera02@gmail.com}
\and
\IEEEauthorblockN{Srividya Subramanian\S{}
\thanks{\S{}Work completed when the author was an exchange student at EPFL.}}
\IEEEauthorblockA{\textit{ETHZ; EPFL} \\
srividya.ssa@gmail.com}
\linebreakand
\IEEEauthorblockN{Thaqiya Aman\P{}
\thanks{\P{}Work completed when the author was an intern at EPFL;
currently at the University of Galway, Ireland.}}
\IEEEauthorblockA{\textit{Presidency University Bangalore; EPFL} \\
thaqiyaaman@gmail.com}
\and
\IEEEauthorblockN{Sara Vaccino}
\IEEEauthorblockA{\textit{EPFL} \\
sara.vaccino@epfl.ch}
\and
\IEEEauthorblockN{Qiang Liu\ddag{}\thanks{\ddag{}Corresponding Author}}
\IEEEauthorblockA{\textit{EPFL}\\
cyruscyliu@gmail.com}
\and
\IEEEauthorblockN{Mathias Payer}
\IEEEauthorblockA{\textit{EPFL} \\
mathias.payer@nebelwelt.net}
}

\maketitle

\begin{abstract}

\sys is an open-source and ground-truth fuzzing benchmark that enables uniform
fuzzer evaluation and comparison.
\sys was originally released with a research paper published at ACM SIGMETRICS
2021.
This short paper explains the motivation, the design, and the impact of \sys,
with a description of extensions to the original benchmark.

\end{abstract}



\section{Introduction}

Fuzzing is a widely-used dynamic bug discovery technique that feeds randomly
generated inputs to a target program to trigger bugs, which is successful in
finding bugs in open-source~\cite{ossfuzz} and commercial
off-the-shelf~\cite{adobefuzz, msftfuzz, gfuzzforsec} software.
This success has resulted in an explosion of new techniques claiming to improve
bug-finding performance~\cite{fuzzingart}.
To highlight improvements, these techniques are typically evaluated across a
range of metrics, including:
\begin{inlineroman} \item crash counts; and/or \item code-coverage profiles.
\end{inlineroman}
While these metrics provide some insight into a fuzzer's performance, they are
insufficient for use in fuzzer comparisons.

The simplest fuzzer evaluation method is to count and compare the number of
crashes triggered by fuzzers on the same target. Unfortunately, crash counts
often inflate the number of actual bugs in the target~\cite{fuzzeval}. Moreover,
deduplication techniques (e.g., coverage profiles, stack hashes) are imprecise
because they fail to accurately identify the root cause of these
crashes~\cite{fuzzeval, aurora}.

Code-coverage profiles are another performance metric commonly used to evaluate
and compare fuzzing techniques. Intuitively, covering more code correlates with
finding more bugs. However, previous work~\cite{fuzzeval} has shown that there
is a weak correlation between coverage-deduplicated crashes and ground-truth
bugs, implying that higher coverage does not necessarily indicate better fuzzer
effectiveness.

The deficiencies of existing performance metrics call for a rethinking of
fuzzer evaluation practices. In particular, the performance metrics used in
these evaluations must accurately measure a fuzzer's ability to achieve its main
objective: \emph{finding bugs}.
Furthermore, the targets that are used to assess how well a fuzzer meets this
objective must be realistic and exercise diverse behavior. This allows a
practitioner to have confidence that a given fuzzing technique will yield
improvements in \emph{finding real-world bugs}.

To satisfy these criteria, we present \emph{\sys}, a ground-truth fuzzer
benchmark based on real programs with real bugs.
\sys consists of \emph{\systargetcount} widely-used open-source libraries and
applications, totaling \SI{2}{\mega\loc} (see~\autoref{sec:target-selection}).
For each \sys program, we manually analyze security-relevant bug reports and
patches, reinserting defective code back into these \systargetcount programs (in
total,~\emph{\sysbugcount} bugs were analyzed and reinserted)
(see~\autoref{sec:bug-selection}).
Additionally, each reinserted bug is accompanied by a light-weight \emph{oracle}
that detects and reports if the bug is \emph{reached} (i.e., the bug code
location is executed) or \emph{triggered} (i.e., the bug condition is satisfied)
(see~\autoref{sec:performance-metrics}).
This distinction between reaching and triggering a bug—in addition to a fuzzer’s
ability to detect a triggered bug—presents a new opportunity to evaluate a
fuzzer across multiple dimensions (again, focusing on ground-truth bugs).
\sys is open-source and available at \sysurl.

\section{The \sys{} Artifact}

\subsection{Target Selection}
\label{sec:target-selection}

\sys contains \systargetcount targets, which we summarize in
\autoref{tab:target-summary}. In addition to these \systargetcount
\emph{targets} (i.e., the codebases into which bugs are injected), \sys also
includes~\sysdrivercount \emph{drivers} (i.e., executable programs that provide
a command-line interface to the target) that exercise different functionality
within the target. Inspired by Google OSS-Fuzz~\cite{ossfuzz}, these drivers are
sourced from the original target codebases (as drivers are best developed by
domain experts).

\begin{table*}[t]
\centering
\caption{The targets, driver programs, bug counts, and evaluated features
incorporated into \sys{}.}
\label{tab:target-summary}
\footnotesize

  \rowcolors{2}{white}{rowgray}
  \begin{tabular}{l>{\raggedright}m{35mm}m{10mm}lrC{10mm}C{10mm}ccC{10mm}}
  \toprule
  Target              & Drivers                                                                              & Version              & File Type           & Bugs         & Magic Values & Recursive Parsing & Compression & Checksums & Global State \\
  \midrule
  \textit{libpng}     & \texttt{read\_fuzzer}, \texttt{readpng}                                              & 1.6.38               & PNG                 & \libpngbugn  & \cmark       & \xmark            & \cmark      & \cmark    & \xmark       \\
  \textit{libtiff}    & \texttt{read\_rgba\_fuzzer}, \texttt{tiffcp}                                         & 4.1.0                & TIFF                & \libtiffbugn & \cmark       & \xmark            & \cmark      & \xmark    & \xmark       \\
  \textit{libxml2}    & \texttt{read\_memory\_fuzzer}, \texttt{xml\_reader\_for\_file\_fuzzer},
                        \texttt{xmllint}                                                                     & 2.9.10               & XML                 & \libxmlbugn  & \cmark       & \cmark            & \xmark      & \xmark    & \xmark       \\
  \textit{poppler}    & \texttt{pdf\_fuzzer}, \texttt{pdfimages}, \texttt{pdftoppm}                          & 0.88.0               & PDF                 & \popplerbugn & \cmark       & \cmark            & \cmark      & \cmark    & \xmark       \\
  \textit{openssl}    & \texttt{asn1}, \texttt{asn1parse}, \texttt{bignum}, \texttt{bndiv},
                        \texttt{client}, \texttt{cms}, \texttt{conf}, \texttt{crl},
                        \texttt{ct}, \texttt{server}, \texttt{x509}                                          & 3.0.0                & \emph{Binary blobs} & \opensslbugn & \cmark       & \xmark            & \cmark      & \cmark    & \cmark       \\
  \textit{sqlite3}    & \texttt{sqlite3\_fuzz}                                                               & 3.32.0               & SQL queries         & \sqlitebugn  & \cmark       & \cmark            & \xmark      & \xmark    & \cmark       \\
  \textit{php}        & \texttt{exif}, \texttt{json}, \texttt{parser}, \texttt{unserialize}                  & 8.0.0{\textminus}dev & \emph{Various}      & \phpbugn     & \cmark       & \cmark            & \xmark      & \xmark    & \xmark       \\ \bottomrule
  \end{tabular}
\end{table*}

\sys's \systargetcount targets were selected for their diversity in
functionality (summarized in \autoref{tab:target-summary}), allowing \sys to
evaluate the code exploration capabilities of fuzzers.
Inspired by benchmarks in other
fields~\cite{dacapo,renaissance,benchmarksimilarity,programsimilarity}, we apply
\emph{Principal Component Analysis} (PCA) to quantify this diversity.
PCA is a statistical analysis technique that transforms an $N$-dimensional space
into a lower-dimensional space while preserving variance as much as
possible~\cite{pca}.
As shown in~\autoref{fig:target-pca}, unsurprisingly, the four \lavam targets
are tightly clustered over the first four principal components, since the \lavam
targets are all sourced from coreutils and hence share the same codebase.
In contrast, both the CGC and \sys provide a wide-variety of
targets. For example, \textit{openssl}---which contains a large amount of
cryptographic and networking code---appears distinct from the main clusters in
\autoref{fig:target-pca}.

\begin{figure}
  \centering
  \begin{subfigure}{\linewidth}
    \centering
    \includegraphics[width=0.8\linewidth,keepaspectratio]{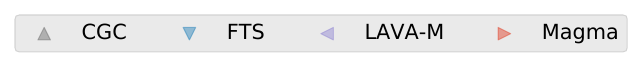}
  \end{subfigure}
  \begin{subfigure}{\linewidth}
    \centering
    \includegraphics[width=\linewidth,keepaspectratio]{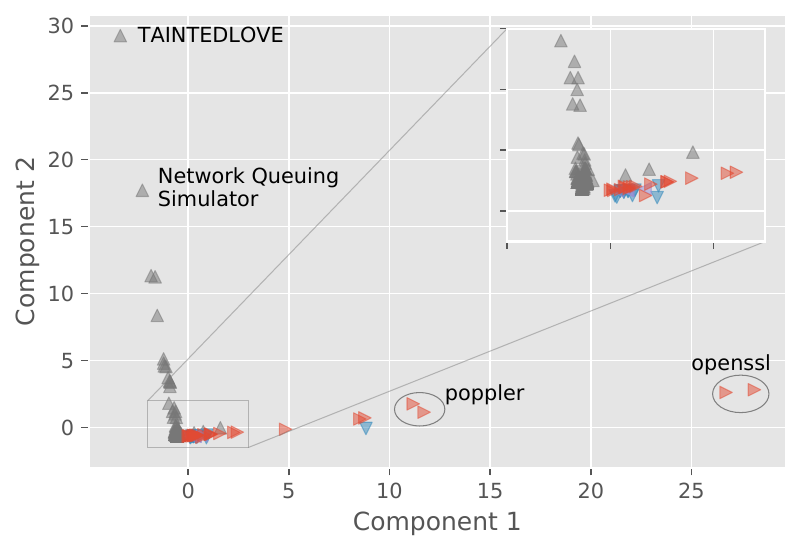}
  \end{subfigure}
  \begin{subfigure}{\linewidth}
    \centering
    \includegraphics[width=\linewidth,keepaspectratio]{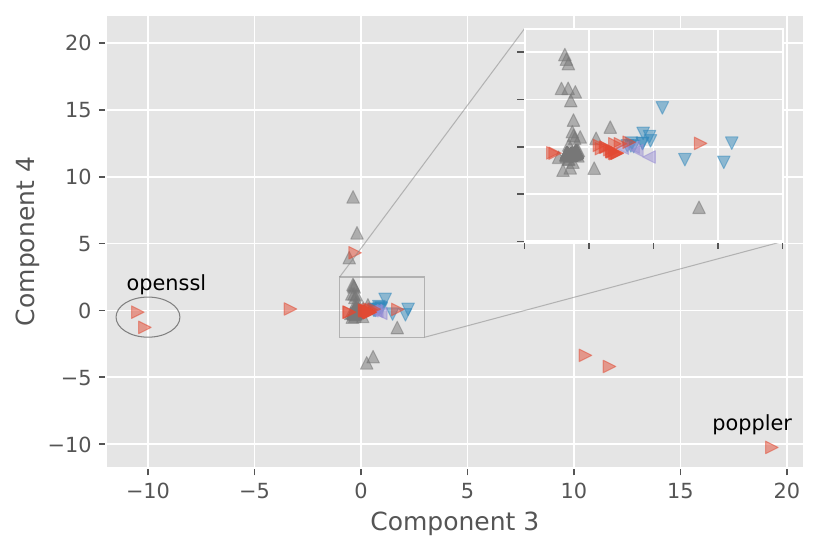}
  \end{subfigure}

  \caption{
  Scatter plots of benchmark scores over the first four principal components
  (~60\% variance in the benchmark targets). Each point represents a benchmark subject and
  closer points indicate lower target diversity.}
  \label{fig:target-pca}
\end{figure}

\subsection{Bug Selection and Insertion}
\label{sec:bug-selection}

\sys contains~\sysbugcount bugs, spanning~\syscwecount CWEs (summarized in
\autoref{fig:bug-summary}). Compared to existing benchmarks, \sys has both the
second-largest variety of bugs (by CWE) and the second-largest ``bug density''
(the ratio of the number of bugs to the number of targets) after the CGC and
\lavam, respectively. While the CGC has a wider variety of bugs, its targets are
not indicative of real-world software (in terms of both size and complexity).
Similarly, while \lavam's bug density (\num{566.25} bugs per target) is an
order-of-magnitude larger than \sys's (\num{16.86} bugs per target), \lavam is
restricted to a single, synthetic bug type.

\begin{figure}
\centering
\includegraphics[width=\linewidth,keepaspectratio]{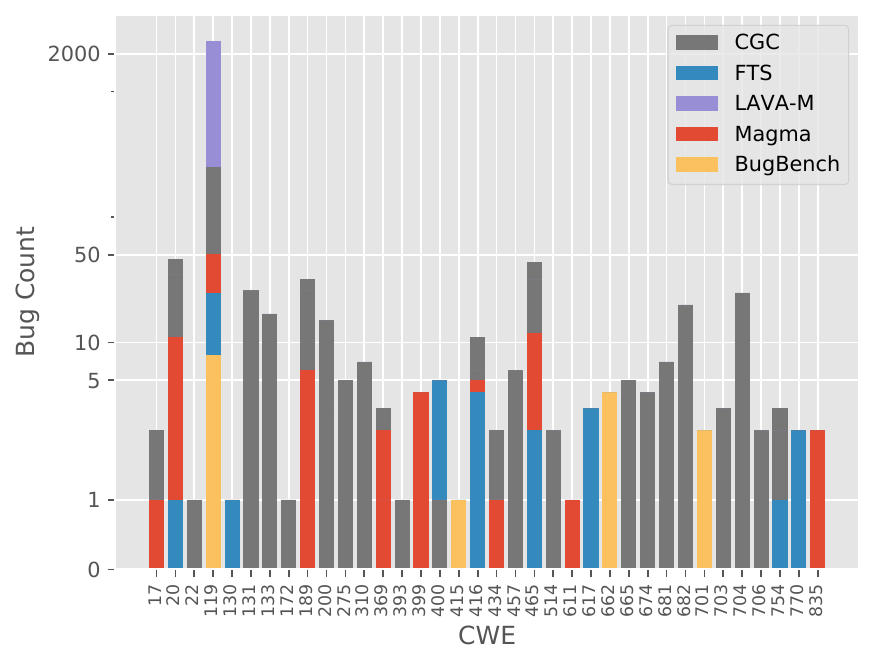}

\caption{Comparison of benchmark bug classes.}
\label{fig:bug-summary}
\end{figure}

Importantly, \sys contains \emph{real} bugs sourced from bug reports and
\emph{forward-ported} to the most recent version of the target codebase. This is
in contrast to existing fuzzing benchmarks (e.g., BugBench, Google FTS) that
rely on old, unpatched versions of the target codebase. Unfortunately, using
older codebases limits the number of bugs available in each target.
In comparison, forward-porting---which is synonymous to back-porting fixes from
newer codebases to older, buggy releases---does not suffer from this issue,
making \sys's targets easily extensible.

Forward-porting begins with the identification---from the reported bug fix---of
the code changes that must be reverted to reintroduce the bug. Bug-fix commits
can contain multiple fixes to one or more bugs, so disambiguation is necessary
to prevent the introduction of unintended bugs. Alternatively, bug fixes may be
spread over multiple commits (e.g., if the original fix did not cover all edge
cases).
Following the identification of code changes, we identify what program state is
involved in evaluating the trigger condition. If necessary, we introduce
additional program variables to access that state. From this state, we determine
a boolean expression that serves as a light-weight oracle for identifying a
triggered bug.  Finally, we identify a point in the program where we inject a
canary before the bug can manifest faulty behavior. This canary helps measure
our fuzzer performance metrics, which are discussed in the following section.

\subsection{Performance Metrics}
\label{sec:performance-metrics}

\sys uses three bug-centric performance metrics, i.e., \emph{reaching},
\emph{triggering}, and \emph{detecting}, to evaluate fuzzers.

A \emph{reached} bug refers to a bug whose oracle was called, implying that the
executed path reaches the context of the bug, without necessarily triggering the
fault. This is where coverage profiles fall short: simply covering the faulty
code does not mean that the program is in the correct state to trigger the bug.
Hence, a \emph{triggered} bug refers to a bug that was reached, and \emph{whose
triggering condition was satisfied}, indicating that a fault occurred.
Source-code instrumentation (i.e., the canary) provides ground-truth knowledge
and runtime feedback of reached and triggered bugs. Each bug is approximated by
\begin{inlinealph}
\item the lines of code patched in response to a bug report, and

\item a boolean expression representing the bug's trigger condition.
\end{inlinealph}
The canary reports:
\begin{inlineroman}
\item when the line of code is reached; and

\item when the input satisfies the conditions for faulty behavior (i.e.,
triggers the bug).
\end{inlineroman}

When a bug is triggered, the oracle only indicates that the conditions for a
fault have been satisfied, but this does not imply that the fault was
encountered or detected by the fuzzer.
Therefore, we also draw a distinction between \emph{triggering} and
\emph{detecting} a bug. Whereas most security-critical bugs manifest as a
low-level security policy violation for which state-of-the-art sanitizers are
well-suited (e.g., memory corruption, data races, invalid arithmetic), other bug
classes are not as easily observed. For example, resource exhaustion bugs are
often detected long after the fault has manifested, either through a timeout or
an out-of-memory error. Even more obscure are semantic bugs, whose malfunctions
cannot be observed without a specification or reference. Consequently, various
fuzzing techniques have been developed to target these bug classes (e.g.,
SlowFuzz~\cite{slowfuzz} and NEZHA~\cite{nezha}). Such advancements in fuzzer
techniques may benefit from an evaluation that includes the bug
\emph{detection} rate as another dimension for comparison.

\subsection{Runtime Monitoring and Post-Processing}
\label{sec:runtime-monitor}

\sys provides a runtime monitor that collects real-time statistics from the
instrumented target. This provides a mechanism for visualizing the fuzzer's
progress and its evolution over time, without complicating the instrumentation.

The runtime monitor collects data about reached and triggered bugs.
Because this data primarily relates to the fuzzer's program exploration
capabilities, we post-process the monitor's output to study the fuzzer's fault
detection capabilities.
This is achieved by replaying the crashing inputs (produced by the fuzzer)
against the benchmark canaries to determine which bugs were reached and/or
triggered.
Furthermore, crashing inputs were validated by replaying them through the
ASAN-instrumented targets, enabling us to assess the fuzzer’s bug detection
capability.

\section{Impact}

Magma is actively used and maintained by the research community; until Sept 2025,
it has been cited 286 times, forked 108 times, starred 319 times, and received
78 pull requests.
Specifically, Magma has been actually used in peer-reviewed papers~\cite{
adrian2021seed, 
bernard2022fine, 
stephan2022an, 
beaman2022fuzzing, 
shah2022mc2, 
jauernig2022darwin, 
srivastava2022one, 
canakci2022targetfuzz, 
koike2022slopt, 
zhou2022no, 
lee2023learning, 
herrera2023dataflow, 
klooster2023continuous, 
yang20231dfuzz, 
li2023accelerating, 
zhang2023shapfuzz, 
chen2023mufuzz, 
liu2023dsfuzz, 
feng2023sizzler, 
zhang2023profile, 
nikhil2023precise, 
zhou2023practical, 
he2023rltg, 
rong2024valkyrie, 
luo2024make, 
she2024fox, 
xu2024graphuzz, 
xu2024isc4dgf, 
wu2024fine, 
huang2024titan, 
li2024sdfuzz, 
qian2024dipri, 
huang2024everything, 
kukucka2024empirical, 
hao2024giantsan, 
haoran2024ddgf, 
rong2024toward, 
bowen2024siro, 
alexandru2024clog, 
benahmed2024modula, 
xu2024vischeduler, 
shunkai2024better, 
zhang2025low, 
qian2025funfuzz, 
xiao2025robust, 
bao2025alarms, 
zuo2025directed, 
geretto2025libaflgo, 
yu2025xfuzz, 
zheng2025mendelfuzz, 
chen2025critical, 
mo2025rcfuzzer, 
zhou2025kraken, 
ELAHI2024120142, 
haruki2025acecov, 
dimitri2025rosa, 
kieum2025refining, 
dekang2025ma, 
hao2025efficient, 
ling2025giantsan, 
li2025gdfuzz, 
wang2025practical, 
riom2025eval}. 
Next, we will explain how Magma serves as a fuzzing benchmark year by year,
facilitating and continuously pushing the relevant research moving forward.

As shown in~\autoref{tab:my-table}, since Magma's release in 2021, at least 45
high-quality papers have used Magma in their own research.
In 2024 and 2025, the number of papers remained stable at 14.
Notably, these papers were published in well-known security and software
engineering conferences, and even system and programming language conferences,
demonstrating Magma's broad impact.

Magma has been widely adopted to advance various aspects of fuzzing, including
input generation, seed selection and mutation, mutation scheduling, coverage
feedback, and bug sanitization. It has also been used in directed fuzzing and
post-fuzzing analyses, such as crash deduplication.
Moreover, researchers have leveraged Magma to evaluate the acceleration and
practical deployment of fuzzing. Beyond fuzzing, Magma has been applied to
broader tasks such as program analysis, program transformation, and backdoor
detection. Collectively, these applications demonstrate Magma’s impact across
the relevant research community.

\begin{table}[]
\centering
\footnotesize
\caption{}
\label{tab:my-table}
\resizebox{\linewidth}{!}{%
\begin{tabular}{@{}lllll@{}}
\toprule
Year & Venue    & Paper                         & Task                          & Category \\
\midrule
2021 & ISSTA    & \cite{adrian2021seed}         & Seed Selection                & Fuzzing \\
\midrule
2022 & CCS      & \cite{shah2022mc2}            & Directed Fuzzing              & Fuzzing \\
2022 & NDSS     & \cite{jauernig2022darwin}     & Mutation Scheduling           & Fuzzing \\
2022 & ACSAC    & \cite{srivastava2022one}      & Directed Fuzzing              & Fuzzing \\
2022 & ACSAC    & \cite{koike2022slopt}         & Mutation Scheduling           & Fuzzing \\
2022 & ASIACCS  & \cite{canakci2022targetfuzz}  & Directed Fuzzing              & Fuzzing \\
2022 & ISSTA    & \cite{bernard2022fine}        & Coverage Feedback             & Fuzzing \\
\midrule
2023 & ICSE     & \cite{lee2023learning}        & Mutation Scheduling           & Fuzzing \\
2023 & TOSEM    & \cite{herrera2023dataflow}    & Coverage Feedback             & Fuzzing \\
2023 & SBFT     & \cite{klooster2023continuous} & Fuzzing Deployment            & Deployment \\
2023 & ISSTA    & \cite{yang20231dfuzz}         & Directed Fuzzing              & Fuzzing \\
2023 & OOPSLA   & \cite{li2023accelerating}     & Coverage Feedback             & Fuzzing \\
2023 & NDSS     & \cite{zhang2023shapfuzz}      & Byte Selection and Mutation   & Fuzzing \\
2023 & Security & \cite{chen2023mufuzz}         & Parallel Fuzzing              & Acceleration \\
2023 & CCS      & \cite{liu2023dsfuzz}          & Input Generation              & Fuzzing \\
2023 & CCS      & \cite{zhang2023profile}       & Fuzzing Acceleration          & Acceleration \\
2023 & ASE      & \cite{nikhil2023precise}      & Fuzzing for Program Analysis  & Others \\
\midrule
2024 & ASIACCS  & \cite{luo2024make}            & Seed Selection                & Fuzzing \\
2024 & CCS      & \cite{she2024fox}             & Seed Selection and Mutation   & Fuzzing \\
2024 & TOSEM    & \cite{xu2024graphuzz}         & Seed Selection                & Fuzzing \\
2024 & TOSEM    & \cite{wu2024fine}             & Coverage Feedback             & Fuzzing \\
2024 & TOSEM   & \cite{qian2024dipri}          & Seed Selection                & Fuzzing \\
2024 & S\&P     & \cite{huang2024titan}         & Directed Fuzzing              & Fuzzing \\
2024 & S\&P     & \cite{huang2024everything}    & Directed Fuzzing              & Fuzzing \\
2024 & Security & \cite{li2024sdfuzz}           & Directed Fuzzing              & Fuzzing \\
2024 & Security & \cite{rong2024toward}         & Directed Fuzzing              & Fuzzing \\
2024 & ISSTA    & \cite{kukucka2024empirical}   & Mutation Scheduling           & Fuzzing \\
2024 & ISSTA    & \cite{haoran2024ddgf}         & Directed Fuzzing              & Fuzzing \\
2024 & ASPLOS   & \cite{hao2024giantsan}        & Bug Sanitization              & Fuzzing \\
2024 & ASPLOS   & \cite{bowen2024siro}          & Program Transformation        & Others \\
2024 & TSE      & \cite{shunkai2024better}      & Seed Selection and Mutation   & Fuzzing \\
\midrule
2025 & TOSEM    & \cite{qian2025funfuzz}        & Coverage Feedback             & Fuzzing \\
2025 & TOSEM    & \cite{hao2025efficient}       & Coverage Feedback             & Fuzzing \\
2025 & Security & \cite{zhang2025low}           & Input Generation              & Fuzzing \\
2025 & Security & \cite{xiao2025robust}         & Coverage Feedback             & Fuzzing \\
2025 & Security & \cite{bao2025alarms}          & Directed Fuzzing              & Fuzzing \\
2025 & EuroS\&P & \cite{geretto2025libaflgo}    & Directed Fuzzing              & Fuzzing \\
2025 & EuroS\&P & \cite{haruki2025acecov}       & Coverage Feedback             & Fuzing \\
2025 & ISSTA    & \cite{yu2025xfuzz}            & Modular-Based Fuzzing         & Implementation \\
2025 & ISSTA    & \cite{zhou2025kraken}         & Parallel Fuzzing              & Acceleration \\
2025 & FSE      & \cite{zheng2025mendelfuzz}    & Mutation Scheduling           & Fuzzing \\
2025 & FSE      & \cite{kieum2025refining}      & Crash Deduplication           & Post-Fuzzing \\
2025 & ICSE     & \cite{chen2025critical}       & Directed Fuzzing              & Fuzzing \\
2025 & ICSE     & \cite{dimitri2025rosa}        & Fuzzing for Backdoor Detection& Others \\
2025 & ICSE     & \cite{wang2025practical}      & Bug Sanitization              & Fuzzing \\
\bottomrule
\end{tabular}%
}
\end{table}

\section{Magma v1.3}

Magma v1.0 (2020) was initially designed with seven targets and 118 ground-truth
bugs. It was soon updated to v1.2 (2021), which remained stable with nine
targets and a total of 138 ground-truth bugs.
However, keeping Magma up to date is essential to staying relevant.
We found that 57 out of 138 bugs in the old version of Magma could no longer be
applied due to code changes.
We also searched public CVE databases and found 246 new CVEs affecting Magma’s
targets since Magma v1.2.
Without such update, Magma would no longer work with the latest fuzzers or
reflect the vulnerabilities that are being discovered recently.

In 2025, we update Magma from version v1.2 to v1.3.
Our updates include: bringing targets and fuzzers to recent versions, fixing
broken bug patches so they work with the new versions (11 bug patches going to
the graveyard), and improving the infrastructure to make it easier to debug and
validate fuzzing results. In addition to this, we introduce tools for automated
versioning and patching. We also add a Proof of Concept (PoC) mode that can be
used to build a centralized collection of bug-triggering inputs and crash logs,
giving users stronger evidence that a fuzzer worked correctly.
This update to v1.3 shows that with structured updates and thoughtful tooling,
Magma can continue to serve as a practical and realistic benchmark for
evaluating fuzzers on real bugs in modern software.

To test our updates, we ran 24-hour fuzzing campaigns using
AFL++~\cite{aflplusplus}, Honggfuzz~\cite{honggfuzz}, and
libFuzzer~\cite{libfuzzer} on nine targets. AFL++ triggered the most bugs (40
total), followed by Honggfuzz (28 total)
and libFuzzer (10 total). Of the 127 bugs in the benchmark, 77 were
reached during fuzzing and 43 were triggered. Among the 57 bugs that were ported
from older versions, 91\% were reached and 34\% were successfully triggered.
These results show that the updated bugs are still meaningful and reachable, and
that the benchmark provides useful data on fuzzer performance.

This update turns Magma into a sustainable benchmark that can evolve with the
fuzzing ecosystem. We introduce automated tools for versioning and patching,
update the infrastructure to support modern environments, and add new features
for debugging and bug validation. Together, these changes make Magma v1.3 a
reliable and future-proof benchmark for fuzzing research.

\section{Conclusion and Outlook}

\sys enables accurate and consistent fuzzer evaluation and performance
comparison with an open ground-truth fuzzing benchmark, where we
forward-ported~\sysbugcount bugs across \systargetcount diverse targets (Magma
v1.0).
This was followed by Magma v1.2, which includes nine targets and 138
ground-truth bugs.
\sys has gained community recognition, and we have continued to improve Magma,
releasing v1.3 in 2025,with nine updated targets and 127 ground-truth bugs.
By updating the software, refining the infrastructure, and making bugs easier to
test, Magma v1.3 advances the state of fuzzing benchmarks. It offers researchers
an up-to-date, realistic, and repeatable way to test fuzzers in a ground-truth
setting.
More importantly, it provides a path forward for maintaining such a benchmark,
ensuring that the comparisons it provides remain relevant in the future.

\printbibliography

\end{document}